# Powerful and Tunable THz Emitters Based on the Fe/Pt Magnetic Heterostructure

Dewang Yang,[1,2#] Jianhui Liang,[3#] Chao Zhou[3], Lu Sun[3], Ronger Zheng[2], Shengnian Luo[1,4], Yizheng Wu[3*], and Jingbo Qi[1,4*]

[1]The Peac Institute of Multiscale Sciences, Chengdu, Sichuan 610031, People's Republic of China

[2]Department of Physics, Optics and Optoelectronics Laboratory, Ocean University of China, Qingdao, Shandong 266100, People's Republic of China

[3]Department of Physics, State Key Laboratory of Surface Physics and Collaborative Innovation Center of Advanced Microstructures, Fudan University, Shanghai 200433, People's Republic of China

[4]Key Laboratory of Advanced Technologies of Materials, Ministry of Education, Southwest Jiaotong University, Chengdu, Sichuan 610031, People's Republic of China

[#] These authors contributed equally to this work.
*Corresponding Authors: **jqi@pims.ac.cn** or **wuyizheng@fudan.edu.cn**

Terahertz (THz) wave, which lies in the frequency gap between infrared and microwave, has an electromagnetic spectrum conventionally defined in the range from 0.1 to 30 THz. [1–3] Because its corresponding photon energy has a scale of milli-electron volt (meV) coinciding with the energy scale of many collective excitations in materials,[1] it has a great potential in fundamental scientific research,[4–7] THz imaging[8, 9] and security applications[3]. Driven by these scientific and technological prospects, many efforts have thus been directed towards the development of new THz sources which are powerful, broadband, low-cost, and readily controllable. [12] In fact, most broadband THz emitters nowadays are based on the excitation of different materials with femtosecond laser pulses. The mechanisms involved in these THz emitters normally are photo-carrier acceleration in photoconducting antennas, second order nonlinear effect in electro-optical crystals, and plasma oscillations. [2, 10–20]

Very recently, it was reported that THz emission can be realized in heterostructures composed of ferromagnetic (FM) and non-FM metal films upon excitation by ultrafast laser pulses, [21–23] where the inverse spin Hall effect (ISHE) plays the key role. In such experiments, the FM film is magnetized by an in-plane magnetic field. Then the heterostructure is pumped by the femtosecond laser pulses, which lead to the generation of non-equilibrium spin polarized electrons in the FM film. The subsequent superdiffusion of these spin polarized electrons into the neighboring non-FM film results in spin current. [24] Due to the ISHE, the spin current in the metal films converts into a transient transverse charge current, which gives rise an electromagnetic wave in the THz region.[25] The strong correlation between the electromagnetic wave emission and the spin polarization thus provides additional degrees of freedom to control the emitted THz wave. However, up to now related studies are rarely explored. In particular, it is still quite elusive that how to manipulate the THz wave properties in specifically designed materials based on such a magnetic heterostructure.

In this communication, we have carried out a comprehensive investigation of THz emission from Fe/Pt magnetic heterostructures, employing time-domain THz spectroscopy. We reveal that by properly tuning the thickness of Fe or Pt layer, THz emission can be greatly improved in this type of heterostructure. We further demonstrate that the THz field strength emitted from a newly designed multilayer [Pt/Fe/MgO]$_n$ with $n=3$ can reach a value of ~1.6 kV/cm, which has the same order as those from the GaAs antenna and the nonlinear crystals, e.g., 100 μm GaP and 2 mm ZnTe, available in the commercial market. Polarization of the emitted THz wave has been shown to follow the rotation of the applied magnetic field. For the first time, the intensity and spectrum of THz wave is demonstrated to be tunable by the magnetic field applied on the patterned magnetic Fe/Pt heterostructures. These findings thus promise novel approaches to fabricate powerful and tunable THz emitters.

**Figure 1**(a) schematically shows the THz wave emission from our Fe/Pt heterostructure under photoexcitation. The FM Fe layer is first magnetized along the *x*-axis. Femtosecond laser pulses then excite electrons in both Fe and Pt films from valence band to conduction band. Because the electron density, band velocity and lifetime of the spin-up (majority) electrons are drastically different from those of the spin-down (minority) electrons in the Fe film, [21] a strong superdiffusive spin current **J**$_s$ will be generated towards the *z* direction with its spin polarization parallel to the magnetization (**M**) direction. [26] This spin current **J**$_s$ enters the Pt film and is converted into a transient transverse charge current **J**$_c$ along the y direction by the ISHE. As a result, an electromagnetic wave with a THz frequency is produced by this transient charge current. In principle, THz wave emission will appear in both +*z* and -*z* directions. Here we only study the former case given the experimental configuration for signal detection. Figure. 1(b) shows the typical time-domain signal $E_{THz}(t)$ in a Pt (5 nm)/Fe (1 nm) sample. The corresponding frequency-domain spectrum evidently illustrates that the emitted electromagnetic wave resides in a bandwidth from ~0.1 to ~4 THz. We have also repeated the same measurements on Co/Pt and Ni/Pt bilayers. Our results indicate that the Fe/Pt bilayer delivers the strongest THz wave emission (see the Supporting Information). Therefore, in the later discussion we only focus on the Fe/Pt bilayer and its derived structures.

In order to obtain the optimized THz wave emission, we have investigated in detail the thickness dependence of $E_{THz}(t)$ in the Fe/Pt bilayer samples. Therefore, we fabricated two types of magnetic heterostructures. One consists of a 3 nm Pt film and a wedged Fe layer, with the Fe layer thickness continuously varying between 0 and 4.8 nm. The other is made of a 1.4 nm thick Fe layer and a wedged Pt film, with the Pt layer thickness ranging from 0 to 5.5 nm. Both wedged samples were fabricated under the same growth conditions. Figure. 1(d)-(e) show the peak field $(E_{THz})_{max}$ and the time-domain THz signal $E_{THz}(t)$ as a function of Fe layer thickness. It is evident that both $(E_{TH})_{max}$ and $E_{THz}(t)$ undergo a rapid increase as this thickness changes from 0 to ∼0.6 nm. When the Fe thickness further increases, $(E_{THz})_{max}$ and $E_{THz}(t)$ experience a slight decrease, and then approximately reach a saturated value at thickness larger than ∼3 nm. We thus conclude that the Fe layer thickness for the strongest THz emission is about 0.6 nm. The quick increase in THz signal for the Fe layer thickness below ∼0.6 nm is attributed to the establishment of magnetic order. The long-range FM order in the Fe layer is crucial, since the spin polarization plays the key role in the ISHE-induced THz wave emission. Our magneto-optical Kerr effect (MOKE) data suggest that the Fe film gradually evolves from the paramagnetic phase to a FM phase, which has an in-plane easy axis at a transition thickness of ∼0.4 nm. This value is consistent with the above critical thickness of 0.6 nm, taken into account the uncertainties introduced by the much larger beam spot-size of the excitation light in current experiment. The slight signal weakening for the Fe films thicker than ∼0.6 nm is likely caused by the back scattering and diffusion of excited electrons in the Fe film away from the Fe/Pt interface. We note that the physical mechanism of generating THz wave is the same at different Fe layer thickness because the THz spectrum shape undergoes little changes with varying thicknesses (see Figures. (f)-(h)).

THz signals for the Pt wedge sample are shown in Figures. 1(i)-(j). The signals undergo a fast increase as the thickness of the Pt film increases from 0 to ∼3 nm, and then almost saturate for thickness larger than ∼3 nm. The THz spectrum shape also remained the same when the Pt layer thickness changes (see Figures. (k)-(m)). Such thickness dependence is possibly caused by spin diffusion. For Pt films less than spin diffusion length, $\lambda$, as the film thickness increases, the transverse charge current increases due to the penetration depth of spin current enlarging. For Pt films thicker than $\lambda$, the THz wave emission will not be enhanced by further increasing the Pt layer thickness, since the transverse charge current will not increase owing to the spin current not able to penetrate into the depth at a distance from the Pt/Fe interface larger than $\lambda$. Therefore, the THz wave signals should saturate at the Pt layer thicknesses beyond $\lambda$. The spin diffusion length $\lambda$ in the Pt film is a few nm, [27,28] consistent with our experimental result.

In Ref. [21], the THz emission as a function of total film thickness has been investigated, and discussed by the Fabry-Perot cavity model. Although this model explained well the intensity drop for large film thicknesses, its derived values clearly deviate from the experimental data for the total thickness less than a critical value $d_c$ (~4 nm), which gives the strongest THz emission. However, such deviation might be understood using the results in this work. Clearly, the THz emission has distinct thickness dependence in respect to the Fe and Pt layers based on our experimental data. Importantly, two aspects cannot be omitted when optimizing the THz emission in thin film samples: (1) The film thickness will greatly affect the magnetic properties, and thus influence the spin-polarized current; (2) The transient charge current is only present in portion of the non-FM layer when its thickness is larger than the spin diffusion length.

In order to further improve the THz wave emission, we have designed a [Pt(2

nm)/Fe(1 nm)/MgO(2 nm)]$_n$ ($n \geq 1$) multilayer structure as a THz emitter for the first time. The sample structure is sketched in **Figure. 2**(a). Because the MgO layer is insulating, photoinduced transient spin current from each Fe layer can only flow into its neighboring Pt layer, where the transverse charge current in each Pt layer is then generated by ISHE. Such instantaneous excitation ensures the transverse charge currents in all Pt layers are approximately in phase. Thus, integration of several Pt(2 nm)/Fe(1 nm)/MgO(2 nm) heterostructures, in principle, can boost THz wave emission. Figure. 2(b)-(c) show the detected THz wave signals $E_{THz}(t)$, and the maximum electric fields $(E_{THz})_{max}$ for different repeating periods, $n$, respectively. The sample with $n=3$ provides the strongest THz wave emission, and the field strength is enhanced by a factor of ~1.7 compared to the case of $n=1$. When $n$ is larger than 3, the THz signal gradually decreases as $n$ increases. However, the signal is still stronger than that of a single Fe/Pt bilayer ($n=1$). The saturation of THz wave signals at $n=3$ may arise from the multiple reflections at the interfaces/surfaces. Because the THz wave is believed to experience high reflection at the metal interface, the wave emitted from the first Pt layer may not efficiently transmit through the rest Fe/Pt layers for $n>3$. When $n$ is large, i.e. $n>3$, the absorption of incident light entering from the substrate side will gradually become significant, leading to a decrease in the number of photoexcited electrons in the top few Fe/Pt bilayers next to the air side as $n$ increases. This is the reason why the detected THz wave emission drops for $n>3$. Such effect, in principle, can be compensated by increasing the pump fluence. Figure. 2(d) shows the pump-fluence dependence of $(E_{THz})_{max}$ for the sample with $n=3$. As fluence increases, the field strength grows and approaches to a value of ~1.6 kV/cm. The saturation effect at high fluencies was also reported in Refs. [22-23], and can be attributed to the spin accumulation effect and the laser-induced heating effect. The former limits the density of spin-polarized electrons in non-FM layer at high fluences [22]. The latter results in the conductivity change and the decrease of the magnetization [23, 29].

Nevertheless, the measured field strength from our sample surprisingly outperforms the measured value from a 100 μm thick GaP crystal (Figure. 2(e)), is equivalent to the reported value of ~2 kV/cm from a 0.5 mm-gap GaAs antenna at a bias field of 4 kV/cm [30], and approaches to the reported value of ~9 kV/cm from 2 mm ZnTe crystal [31], under the same pump fluence.

Another advantage of the THz emitter based on magnetic heterostructures is that the polarization of emitted THz wave can be easily tuned with an external magnetic field. As depicted in Figures 1(a) and 2(a), the spin polarization of spin current from the Fe layer into the Pt layer can be readily controlled by rotating the magnetization **M** via an external magnetic field **H**. The direction of the transient charge current due to the ISHE is perpendicular to the magnetization direction. As a result, one can control the polarization of emitted THz wave by only rotating the magnetic field. **Figure 3** demonstrates the polarization control. If we fix the magnetic field along the *x*-axis and rotate the sample azimuthally, the measured THz wave strength is almost constant, independent of the rotating angle (with respect to the *y*-axis). This result suggests that the intensity of THz wave emitted from magnetic heterostructure is independent on the magnetic or crystalline anisotropy in the films. This isotropic behavior is regardless of the magnetization orientation and pump polarization. On the other hand, if the sample is fixed but the magnetic field rotates, the polarization of emitted THz wave rotates and is always perpendicular to the in-plane field direction. In our experiment, the detecting setup is always fixed, and the optimized polarization detection is along the *y*-axis due to the anisotropy of the ZnTe(110) detecting crystal. Thus, a $|\sin\alpha|$-like behavior of the absolute field strength is expected (see Figure. 3(a)), when the field rotates by an angle of $\alpha$, representing the orientation with respect to the *y*-axis. A special case is that by reversing the magnetic field, the generated THz wave experiences a phase shift of $\pi$ (Figure. 3(b)), in consistent with the results in Ref. [22].

It should be pointed out that because the Pt/Fe heterostructure is a thin-film THz emitter, it can be readily patterned. If the transient charge current induced by ultrafast laser pulses is confined laterally, one will expect that the THz wave emission is modified accordingly. For instance, when the heterostructure is patterned as stripes, the direction of the transient charge current can be oriented in a controlled manner either parallel or perpendicular to the stripes by rotating the applied field, as shown in **Figure. 4**(a). In this work, we have fabricated a stripe-patterned Fe/Pt sample, with a stripe width of 5 μm and a spacing of 5 μm, to illustrate the corresponding effect. Figure. 4(d) show the THz wave signals in this patterned emitter for several stripe directions. These signals were measured during azimuthally rotating the sample with the field orientation fixed. The rotation angle θ is defined relative to the field orientation. Clearly, the THz wave signal at θ=0° is nearly 40% smaller than that at θ=90°. In frequency domain, the former signal obviously has a blue shift compared with latter, as shown by the normalized spectrums in Figure. 4(e). The peak frequency changes from ~0.6 THz to ~1.3 THz for θ varying from 90° to 0°. Such anisotropic THz emission can be qualitatively understood by the current confinement effect in the stripes. At θ=0° with the magnetic field along the stripes, the transient transverse charge current flows perpendicular to the stripes, and thus will lead to the charge accumulation at the stripe edges. Such charge accumulation will generate a transient opposite electric field, which in turn results in a decrease and fast change of the initial transient charge current. These effects are finally attributed to the observed intensity variation of the emitted THz wave and the shift of peak frequency. At θ=90° with the field perpendicular to the stripes, there will be no current confinement effect since the transient transverse charge current flows along the stripes. It is well known that conventional patterned metamaterial can be employed to effectively control the THz wave properties, such as strength, centre frequency, polarization and bandwidth.[32,33,34]

Metamaterials comprised of split-ring resonator has been proposed as a new broadband THz source.[35] However, our work demonstrates for the first time that the THz emitter made of the patterned magnetic heterostructure is advantageous in its convenience of manipulating intensity, polarization, and spectrum of the emitted THz wave. These findings thus may inspire further investigation on more complex patterned THz emitter based on Fe/Pt or other similar heterostructures.

In conclusion, the magnetic multilayer comprised of a Fe/Pt heterostructure has been demonstrated as a novel type of powerful, broadband and flexible THz emitter, and is capable of emitting THz waves even stronger than that from the conventional nonlinear optical crystals. In addition, we reveal that the intensity, polarization and spectrum of THz wave emitted from a simple stripe-patterned magnetic heterostructure can be conveniently manipulated by rotating an external magnetic field. A magnetic multilayer and its patterned forms bear great potential to be a new type of powerful and tunable THz sources.

Sample Preparations and Experimental Setups

The Fe/Pt heterostructures were grown by molecular beam epitaxy on single-crystalline $Al_2O_3$(0001) substrates in an ultrahigh vacuum (UHV) chamber with a base pressure of $2\times10^{-10}$ Torr. The double-side polished substrates were ultrasonically cleansed by acetone and deionized water. The substrates then were annealed at ~700 °C for 1 hour inside the UHV chamber. The Fe layers were prepared with a thermal evaporator at room temperature. The Pt and MgO films were grown by pulse laser deposition with a 248 nm KrF excimer laser at room temperature [36]. The in-situ reflection high-energy electron diffraction patterns indicate all the films are single-crystal. The film thickness was determined by the growth rate, which is usually 1-2 Å/min measured by a calibrated quartz thickness monitor for each film. The thickness slope for the Fe or Pt wedge samples is ~0.4 nm/mm. All the samples have been covered with a 5 nm MgO capping layer to protect the Fe layer from oxidation before being taken out from the UHV chamber. The stripe-patterned Fe/Pt sample was prepared through a standard photolithography and $Ar^+$ ion etching process. The magnetic properties were characterized by magneto-optic Kerr effect, which shows all the films can be saturated with a field strength less than 20 mT.

In the time-domain THz spectroscopy experiment, all the samples were kept inside a permanent magnet with the in-plane magnetic field of ~200 mT, which is sufficiently strong to align the magnetization to the field direction. The THz wave emission arises from the photoexcitation by linearly polarized femtosecond laser pulses from a Ti:Sapphire regenerative amplifier (with a duration of ~55 fs, a center wavelength of 800 nm, and a repetition rate of 1 KHz). The laser beam was incident from the substrate side. So it excited the magnetic heterostructures after passing through the substrates. The diameter of excitation light beam was ~3 mm on the uniform samples, while a diameter of ~1 mm was used for the Fe or Pt wedge samples. The typical excitation fluence was ~200 $\mu J/cm^2$. Prior to THz wave detection, the residual transmitted laser beam from the samples was blocked by a ~3 mm Teflon filter. The signals were finally detected by the electro-optic sampling technique, where the probe pulses from the same laser co-propagate with the THz wave through a 0.2 mm-thick electro-optic ZnTe (110) crystal. The THz wave was focused on the crystal with a diameter of ~1.5 mm. The width of probe pulse at the detector is ~70 fs, which limits the response function of the detector with a bandwidth of ~4 THz. All measurements were performed at room temperature in a vacuum system.

Acknowledgements

We would like thank Prof. Lei Zhou for helpful discussions. This project was supported by 1000-Youth-Talents Plan, the National Key Basic Research Program of China (Grant No. 2015CB921401), and National Science Foundation of China (Grants No. 11434003, No. 11474066, No. 11274074).

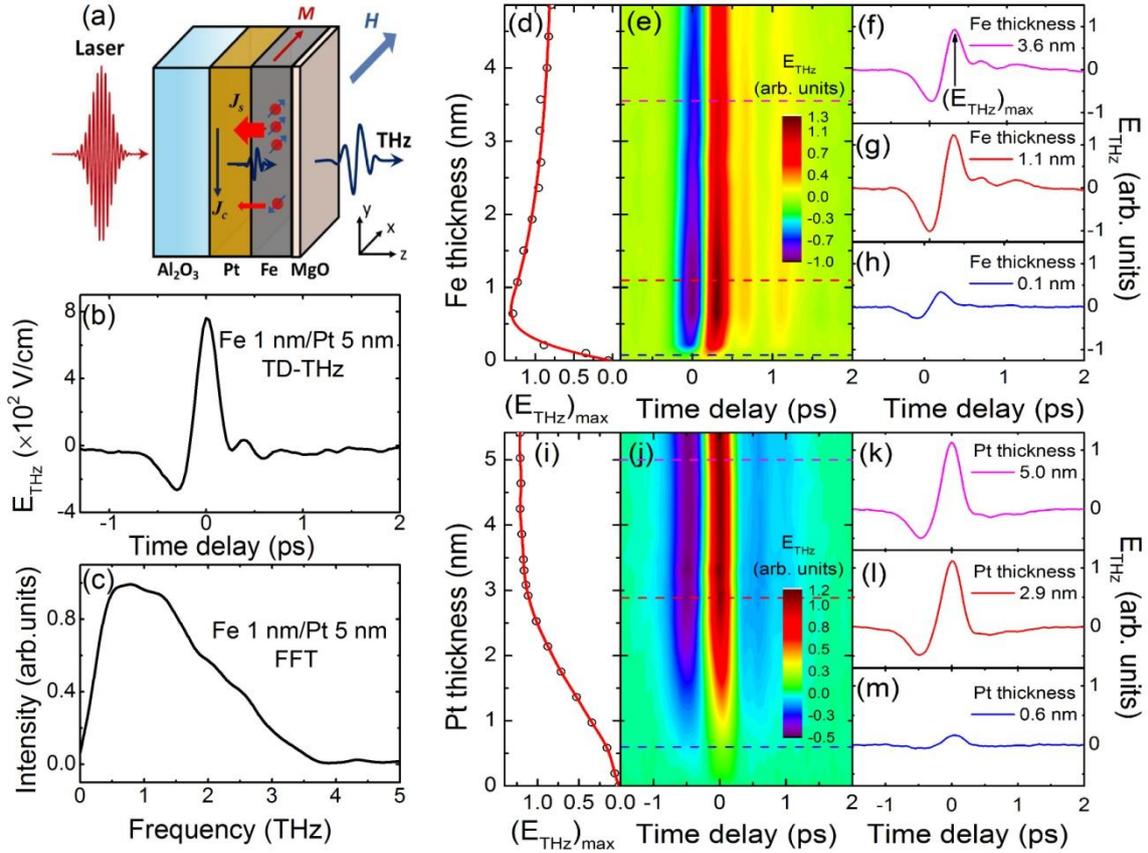

**Figure 1.** (a) Schematic of THz emitter made of the Fe/Pt magnetic heterostructure under an in-plane magnetic field **H** of ∼200 mT. (b) and (c) show typical time- and frequency-domain THz signals from the sample Fe (1 nm)/Pt (5 nm). (d)-(h) are THz wave signals from the sample Fe (wedge)/Pt (3 nm). (d) is the peak field $(E_{THz})_{max}$ as a function of Fe layer thickness. (e) is the 2D false-color plot of the THz temporal traces at different Fe layer thicknesses. (f)-(h) are the time-domain THz signals at three different thicknesses indicated by the corresponding red dashed lines in (e). (i)-(m) are THz signals from the sample Fe (1.4 nm)/Pt (wedge). (i) is the peak field $(E_{THz})_{max}$ as a function of Pt layer thickness. (j) is the 2D false-color plot of the THz temporal traces at different Pt layer thicknesses. (k)-(m) are the time-domain THz signals at three different thicknesses indicated by the corresponding red dashed lines in (j).

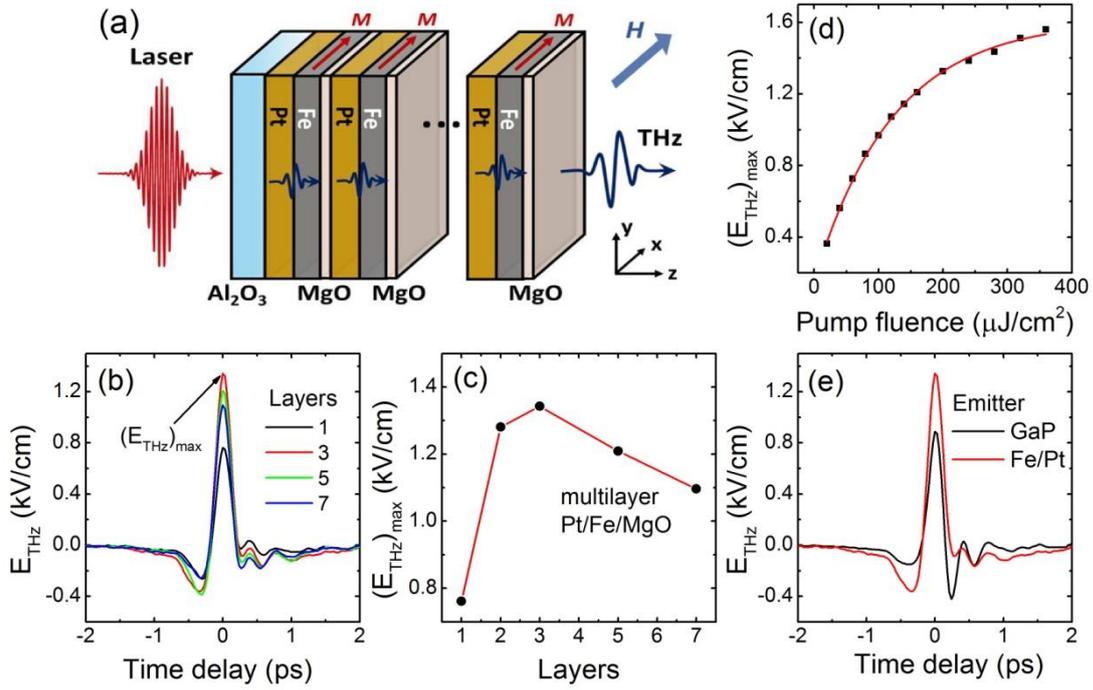

**Figure 2**. (a) Schematic of a THz emitter made of magnetic multilayers. (b) The time-domain THz signals on multilayer structure [Pt(2 nm)/Fe(1 nm)/MgO(2 nm)]$_n$ ($n$=1, 3, 5, 7). (c) The peak field ($E_{THz}$)$_{max}$ of THz wave as a function of $n$. (d) The fluence dependence of ($E_{THz}$)$_{max}$ of the emitter with $n$=3. (e) Comparison between the time-domain THz signal from the multilayer structure with $n$=3 and that from the GaP crystal, at the same pump fluence of 200 μJ/cm$^2$ on the emitters, and the same focused THz wave spot size of ~1.5 mm on the detector ZnTe.

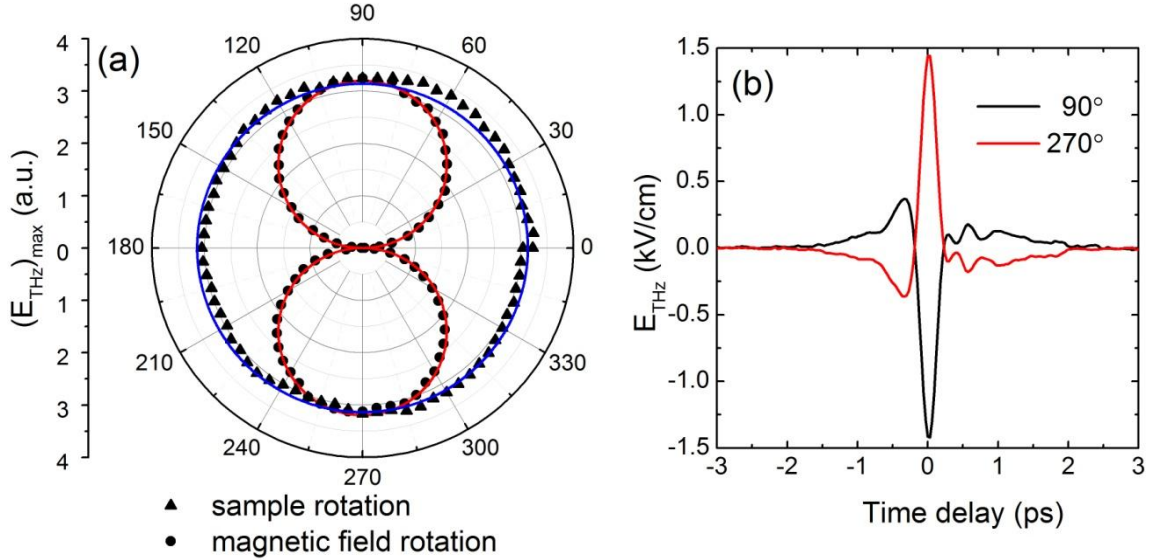

**Figure 3**. (a) The peak field ($E_{THz}$)$_{max}$ of THz wave as a function of the rotation angle α, with respect to the y-axis, either by rotating the magnetic field (circles) or the sample orientation (triangles). The red line is a curve fit proportional to |sinα|. The blue line is a constant fit. (b) The time-domain THz signals with a phase difference of π at two magnetic field directions (90 ° and 270 °).

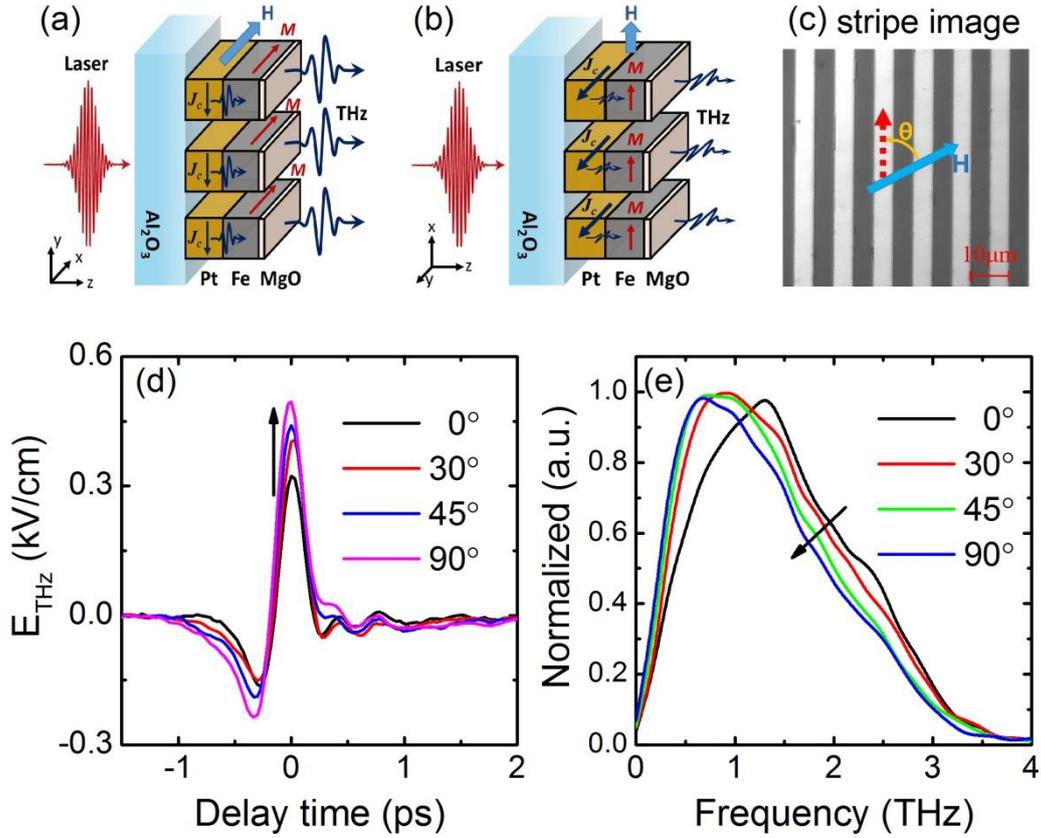

**Figure 4**. Schematic of a patterned magnetic heterostructure with the stripes (a) parallel and (b) perpendicular to the magnetic field. (c) A photograph of the patterned Fe/Pt sample (top view). (d) and (e) show the time- and frequency-domain THz signals at different stripe orientations, respectively. The magnetic field **H** is fixed along +*x* direction in the laboratory coordinate system. The orientation angle θ characterizing the rotation of patterned heterostructure is defined in (c). The black arrows in (d) and (e) represent the angle θ increasing from 0 ° to 90 °.

16